\documentclass{svmult}

\usepackage{makeidx}         
\usepackage{graphicx}        
\usepackage{multicol}        

\begin{document}

\title*{Prediction oriented variant of financial log-periodicity 
        and speculating about  the stock market development until 2010}
\titlerunning{Prediction oriented variant of financial log-periodicity}
\author{Stan Dro\.zd\.z\inst{1,2}, Frank Gr\"ummer\inst{3},
        Franz Ruf\inst{4}\and Josef Speth\inst{3}}
\authorrunning{Stan Dro\.zd\.z et al.}
\institute{Institute of Nuclear Physics, Polish Academy of Sciences,
           PL--31-342 Krak\'ow, Poland
\and Institute of Physics, University of Rzesz\'ow, 
     PL--35-310 Rzesz\'ow, Poland
\and Institut f\"ur Kernphysik, Forschungszentrum J\"ulich,
     D-52425 J\"ulich, Germany
\and West LB International S.A., 32-34 bd Grande-Duchesse Charlotte,
     L-2014 Luxembourg}
%
%
\maketitle

{\bf Summary.} A phenomenon of the financial log-periodicity is discussed
and the characteristics that amplify its predictive potential
are elaborated. The principal one is self-similarity that obeys 
across all the time scales. Furthermore the same preferred scaling factor
appears to provide the most consistent description of the market
dynamics on all these scales both in the bull as well as in the 
bear market phases and is common to all the major markets. These ingredients
set very desirable and useful constraints for understanding the past
market behavior as well as in designing forecasting scenarios. 
One novel speculative example of a more detailed S$\&$P500 development 
until 2010 is presented.\\


\section*{}
The concept of financial log-periodicity is based on the appealing assumption
that the financial dynamics is governed by phenomena analogous to criticality
in the statistical physics sense (Sornette et al. 1996, 
Feigenbaum and Freund 1996). 
Criticality implies a scale invariance which, for a properly defined 
function $F(x)$ characterizing the system, means that  
\begin{equation}
F(\lambda x) = \gamma F(x).
\label{eq:F}
\end{equation}
A constant $\gamma$ describes how the properties of the system 
change when it is rescaled by the factor $\lambda$.
The general solution to this equation reads:
\begin{equation}
F(x) = x^{\alpha} P(\ln(x)/\ln(\lambda)),
\label{eq:FP}
\end{equation}
where the first term represents a standard power-law that is characteristic 
of continuous scale-invariance with the critical exponent 
$\alpha = \ln(\gamma) / \ln(\lambda)$ and $P$ denotes a periodic function
of period one. This general solution can be interpreted in terms of
discrete scale invariance. It is due to the second term that the conventional 
dominating scaling acquires a correction that is periodic in $\ln(x)$ and
may account for the zig-zag character of financial dynamics.  
It demands however that if $x = \vert T - T_c \vert$, where $T$ denotes
the ordinary time labeling the original price time series, represents 
a distance to the critical point $T_c$, the resulting spacing between the
corresponding consecutive repeatable structures at $x_n$ seen in the linear
scale follow a geometric contraction according to the relation
$(x_{n+1}-x_n) / (x_{n+2}-x_{n+1}) = \lambda$.
The critical points correspond to the accumulation of such oscillations and,
in the context of the financial dynamics, it is this effect that potentially 
can be used for prediction. 

An extremely important related element, for 
a proper interpretation and handling of the financial patterns as well as for 
consistency of the theory, is that such log-periodic oscillations manifest
their action self-similarly through various time scales 
(Dro\.zd\.z et al. 1999). This applies both to the log-periodically 
accelerating bubble market phase as well as to the 
log-periodically decelerating anti-bubble phase. Furthermore, more and more 
evidence is collected that the preferred scaling factor $\lambda \approx 2$
and is common to all the scales and markets (Dro\.zd\.z et al. 2003). 
These two elements, 
self-similarity and universality of the $\lambda$, set very valuable
and in fact crucial constraints on possible forms of the analytic 
representations of the market trends and oscillation patterns, 
including the future ones as well. 

A specific form of the periodic function $P$ in Eq.~\ref{eq:FP} is as yet
not provided by any first principles which opens room for certain, seemingly 
mathematically unrigorous assignments of patterns. 
This, on the other hand, allows to correct for frequent market 'imprecisions' 
when relating its real behavior versus the theory. 
Very helpful in this respect is the requirement of self-similarity which
greatly clarifies the significance of a given pattern and allows to determine
on what time scale it operates.  
Since in the corresponding methodology the oscillation structure carries
the most relevant information about the market dynamics, for transparency 
of this presentation, we use the first term of its Fourier expansion,  
\begin{equation}
P(\ln(x)/\ln(\lambda)) = A + B \cos({\omega \over 2\pi} \ln(x) + \phi). 
\label{eq:FPE}
\end{equation}
This implies that $\omega = 2\pi / \ln(\lambda)$. Already such a simple 
parametrization allows to properly reflect the contraction of oscillations,
especially on the larger time scales. On the smaller time scales just 
replacing
the {\it cosine} by its modulus often, even quantitatively in addition,
describes departures of the market amplitude from its average trend.    

\begin{figure}[ht]
\begin{center}
\includegraphics[width=6cm, angle=270]{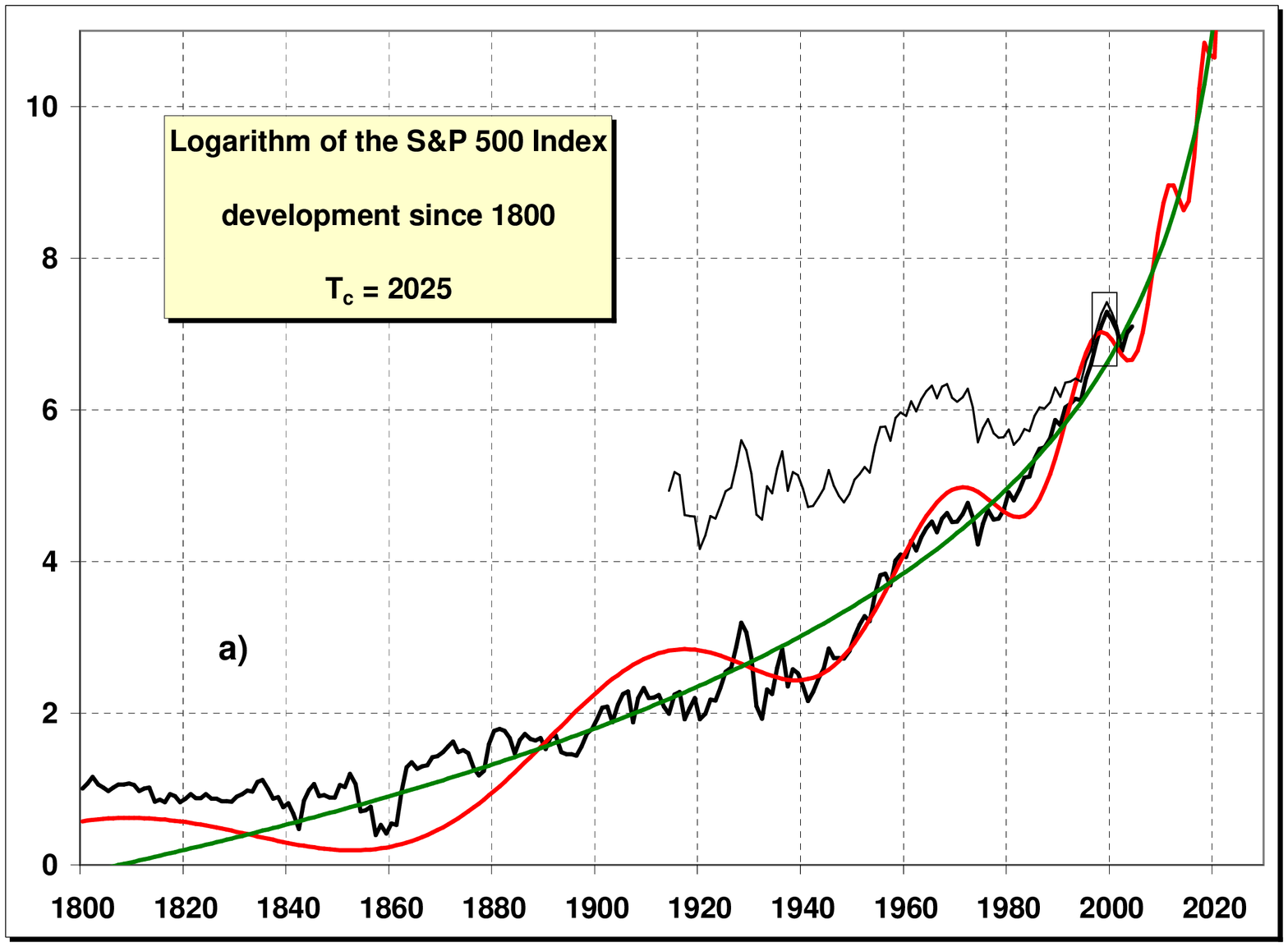}
\includegraphics[width=6cm, angle=270]{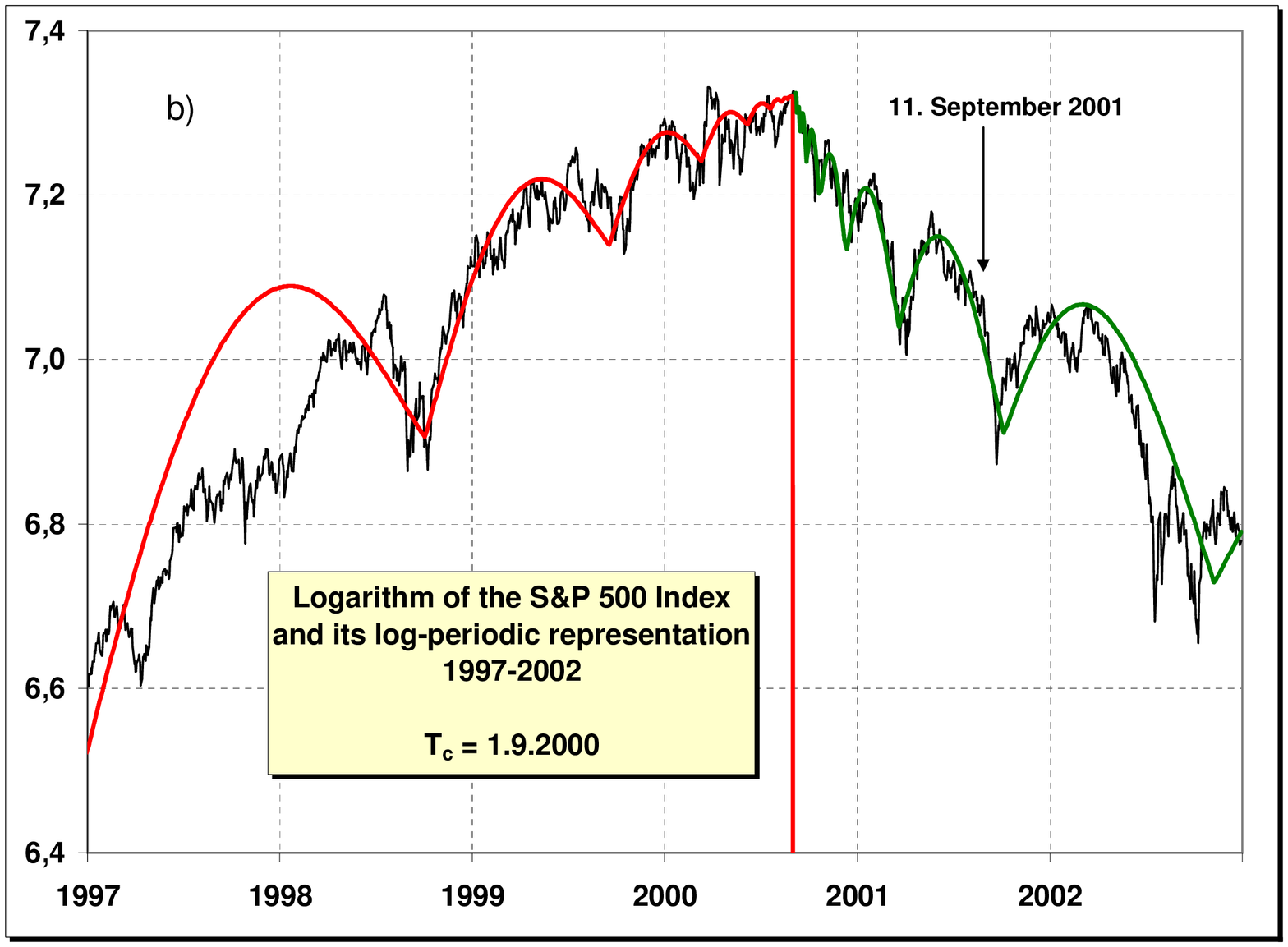}
\caption{(a) Logarithm of the Standard $\&$ Poor's 500 index 
since 1800 (http://www.globalfindata.com). The thick solid line displays its optimal 
log-periodic representation with $\lambda = 2$. The thin solid line 
represents the inflation corrected S$\&$P500 expressed in 2004 US$\$$.   
It significantly shifts the third minimum to the early 1980s and improves 
agreement with the theoretical representation. (b) Logarithm of the S$\&$P500
from 1997 till the end of 2002, which corresponds to the magnification of the
small rectangle in (a). The solid lines illustrate the log-periodic 
accelerating and decelerating representations with $\lambda = 2$, modulus
of the cosine used in Eq.~(3), and a common $T_c = 1.9.2000$.}
\end{center}
\end{figure}

One particularly relevant and special, for several reasons, 
example is shown in Fig.~1. 
The upper panel (a) illustrates a nearly optimal log-periodic representation 
of the S$\&$P500 data over the most extended time-period of the recorded 
stock market activity as dated since 1800. As already pointed out
(Dro\.zd\.z et al. 2003) this development signals in around 2025 a transition 
of the S$\&$P500 to a globally declining phase as measured in the contemporary
terms. The magnification of the small rectangle covering the period
1997-2002 is displayed in the lower panel (b) of the same Fig.~1. It thus 
illustrates the nature of the stock market evolution on a much smaller time
scale of resolution. An impressive log-periodicity with the same $\lambda=2$
on both sides of the transition date (September 1, 2000) can be seen. 

The next stock market top from the perspective of the largest time scale
(Fig.~1a) can be estimated to occur in around the years 2010-2011. 
In the spirit of the log-periodicity its neighborhood is to be accompanied
by the smaller time scale oscillations - similar in character to those in
Fig.~1b.   
Of course, when going far away from those large scale transition points 
such pure log-periodic structures - representative to the one level lower 
time scale - must get dissolved. 
A particularly interesting related question then is what characteristics are 
to
govern the stock market dynamics in the transition period when going from
2000 to 2010. The most natural and straightforward way is to view this process
as schematically is indicated in Fig.~2.          
This whole period is thus covered by the two main components represented by
the thin lines and the market dynamics is driven by the superposition of 
of these two components whose phases, slopes and weights are adjusted such  
that the overall global market trend up to now is reproduced. 

\begin{figure}[ht]
\begin{center}
\includegraphics[width=6cm, angle=270]{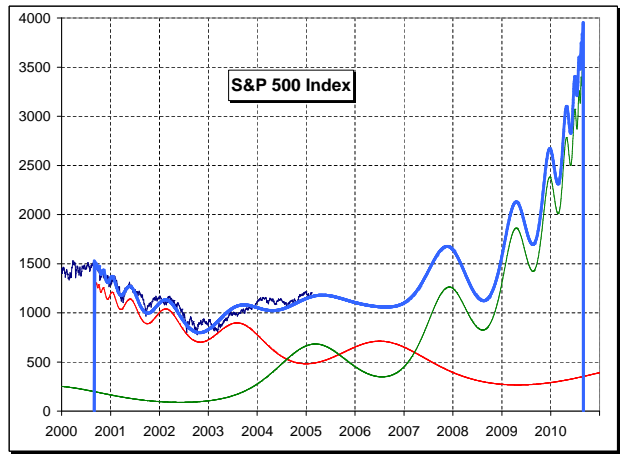}
\caption{A hypothetical log-periodic scenario, 
represented by the thick solid line,
for the S$\&$P500 development until 2010. This solid line is obtained by
summing up the two $\lambda = 2$ components (thin lines): 
log-periodically decelerating since 1.9.2000
and log-periodically accelerating toward 1.9.2010.}    
\end{center}
\end{figure}

In this scenario, close to the two large-scale transition points 
(September 2000 and, as provisionally estimated here based on Fig.~1a, 
September 2010) the market is driven, as needed, 
essentially by the single log-periodic components, 
decelerating and accelerating one, correspondingly.     
More complicated is the situation in the middle of this time interval where
the two components contribute comparably. Most interestingly, it indicates 
that
the period of the stock market stagnation may extend even into the year 2008,
before it seriously starts rising. 
It also demonstrates a possible mechanism that generates modulation structures
responsible for the apparent higher order corrections 
(Johansen and Sornette 1999) to Eq.~(\ref{eq:FPE}).  
The changes in the frequency relations observed in the transition period 
between the bear and the bull market phases originate here from the 
interference between the two components, both of the simple form as prescribed
by Eq.~(\ref{eq:FPE}) and with the same $\lambda = 2$.  
Of course, similar effects of interference may occur on the whole 
hierarchy of different time scales.   

There is one more element that from time to time takes place 
in the financial dynamics and whose identification appears
relevant for a proper interpretation of the financial patterns with
the same universal value of the preferred scaling factor $\lambda$. 
This is the phenomenon of a "super-bubble" (Dro\.zd\.z et al. 2003)
which is a local bubble, itself evolving log-periodically, 
superimposed on top of a long-term bubble. Two such spectacular examples
are provided by the Nasdaq in the first quarter of 2000 and by the gold price
in the beginning of 1981 (Dro\.zd\.z et al. 2003).

\begin{figure}[ht]
\begin{center}
\includegraphics[width=6cm, angle=270]{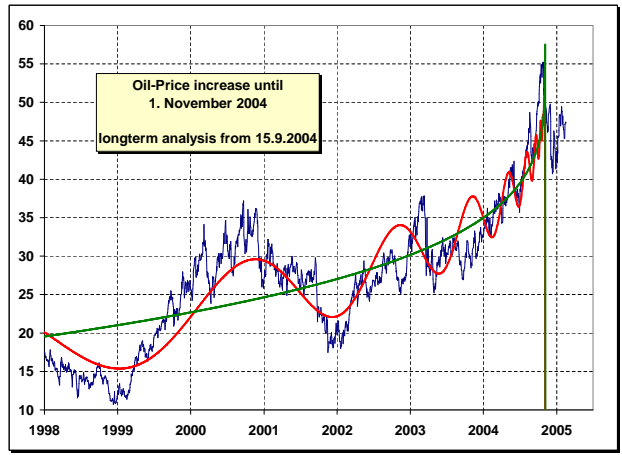}
\caption{The New York traded oil futures since 1998 and the 
corresponding log-periodic $\lambda = 2$ representation in terms of Eq.~(3).}    
\end{center}
\end{figure}

In connection with this second case it is important to remember that the same 
value of $\lambda$ as for the stock market turns out appropriate.  
That such its value may be characteristic to the whole commodities market 
as well,
is shown in Fig.~3 which displays the New York traded oil futures versus
the best log-periodic $(\lambda = 2)$ representation. In fact, this scenario 
has been drawn by the authors on September 15, 2004, insisting on using 
$\lambda = 2$, even though one local minimum (in the beginning of 
2004) in the corresponding sequence did not look very convincing.  
Designed this way it was indicating a continuation of the increase 
until the end of October and then a more serious reverse of the trend.
Subsequent development of the oil futures provides further arguments in favor
of this way of handling the financial log-periodicity.             

\section*{References}

\begin{description}
\item[]Dro\.zd\.z S, Ruf F, Speth J, W\'ojcik M (1999)~
Imprints of log-periodic self-similarity in the stock market.
Eur. Phys. J. B 10:589-593

\item[]Dro\.zd\.z S, Gr\"ummer F, Ruf F, Speth J (2003)~
Log-periodic self-similarity: an emerging financial law?
Physica A 324:174-182

\item[]Feigenbaum JA, Freund PGO (1996)~
Discrete scale invariance in stock markets before crashes.
Int. J. Mod. Phys. B 10:3737-3745

\item[]Johansen A, Sornette D (1999)~
Financial "anti-bubbles": Log-periodicity in gold and Nikkei collapses.
Int. J. Mod. Phys. C 10:563-575

\item[]Sornette D, Johansen A, Bouchaud J.-P (1996)~
Stock market crashes, precursors and replicas.
J. Physique (France) 6:167-175

\end{description}

\end{document}